# Surface Passivation Suppresses Local Ion Motion in Halide Perovskites

*Justin Pothoof, Robert J. E. Westbrook, Rajiv Giridharagopal, Madeleine D. Breshears, and David S. Ginger*

Department of Chemistry, University of Washington, Seattle, Washington, 98195-1700, United States

Corresponding Author: dginger@uw.edu

Abstract

We use scanning probe microscopy to study ion migration in the formamidinium (FA)-containing halide perovskite semiconductor $Cs_{0.22}FA_{0.78}Pb(I_{0.85}Br_{0.15})_3$ in the presence and absence of chemical surface passivation. We measure the evolving contact potential difference (CPD) using scanning Kelvin probe microscopy (SKPM) following voltage poling. We find that ion migration leads to a ~100 mV shift in the CPD of control films after poling with 3V for only a few seconds. Moreover, we find that ion migration is heterogeneous, with domain interfaces leading to a larger shift in the CPD. Application of (3-aminopropyl)trimethoxysilane (APTMS) as a surface passivator further leads to 5-fold reduction in the CPD shift from ~100 meV to ~20 meV. We use hyperspectral microscopy to show that APTMS-treated perovskite films undergo less photoinduced halide migration than control films. We interpret these results as due to a reduction in halide vacancy concentration due to passivation with APTMS.

TOC Graphic

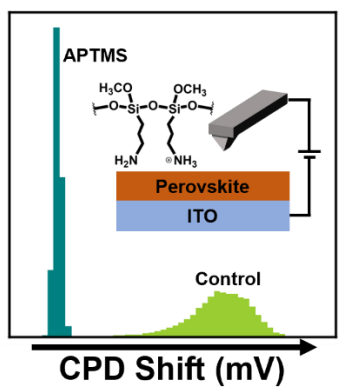

Halide perovskite semiconductors are important materials for a range of optoelectronic applications such as photovoltaics and light-emitting devices,[1,2] and power conversion efficiencies for single-junction perovskite solar cells have increased to over 25.7%.[3] One challenge facing some applications is that perovskites are prone to undergoing ion migration, in which ions move through the crystal lattice.[4,5] Ion migration can lead to undesired interactions between the perovskite active layer and transport layers or electrodes, reduce operational stability, and lead to segregation of the perovskite into separate phases.[6–11] Due to a low energetic barrier, vacancy-mediated halide migration has been proposed to be the dominant ion migration process in the perovskite lattice.[5,12] Many studies have been performed using chemical passivation strategies as a method of reducing surface halide vacancies, which has resulted in increased photoluminescence quantum yields and ultimately device performance.[13–20] However, there are far fewer studies that examine how surface passivation affects ion migration by means other than hysteresis reduction,[21–24] especially at the local level.

Previous work from our group has shown that (3-aminopropyl)trimethoxysilane (APTMS) significantly reduces nonradiative recombination in halide perovskite semiconductors.[25,26] Since halide migration often involves halide vacancies,[5,12] the same sites that are often targeted by chemical surface passivation,[27,28] we hypothesize that surface passivation with molecules such as APTMS should also suppress halide migration.

Here, we examine this hypothesis, with a particular emphasis on investigating how APTMS surface passivation can affect ion migration *locally*, via scanning Kelvin probe microscopy (SKPM). We combine SKPM measurements on locally-poled perovskite samples with studies of photoluminescence using hyperspectral optical microscopy. The use of SKPM allows us to probe ion motion and effects of APTMS surface passivation below the optical diffraction limit of conventional photoluminescence measurements.[29–31] We focus our study on wide-gap (~1.7 eV), mixed-halide perovskites because such formulations are particularly relevant for perovskite-on-Si tandem photovoltaics and because ion migration often causes halide phase segregation in these compositions. We find that the contact potential difference (CPD) of the perovskite samples evolves with applied electric fields. We quantify the average shift in CPD for perovskite control films to be near ~100 mV at poling extremes of 3 V with a poling dwell time of only a few seconds, which is reduced to ~20 mV after surface passivation with APTMS. We attribute this reduction in CPD shift to the passivation of surface halide defects. Using photoluminescence hyperspectral imaging, we also observe a reduction in photoinduced halide segregation in the perovskite films after surface passivation with APTMS.

For this study, we prepared mixed-halide perovskite semiconductor films of the composition $Cs_{0.22}FA_{0.78}Pb(I_{0.85}Br_{0.15})_3$ on ITO substrates by using a one-step spin-coating technique as adapted from the literature[32] (see SI for full details). We refer to these as-grown films as "control" or "unpassivated" perovskites. To prepare passivated perovskite films, we exposed the films to APTMS for five minutes at room temperature in a low-vacuum chamber as previously described.[25,26] We verified the perovskite composition and structure using XRD, and the bandgap using UV-Vis absorption spectroscopy (Figure S1). Figure 1 shows that APTMS passivation lengthens the photoluminescence (PL) lifetimes and increases the PL quantum yield of APTMS-treated samples, as is consistent with a reduction in surface trap states.

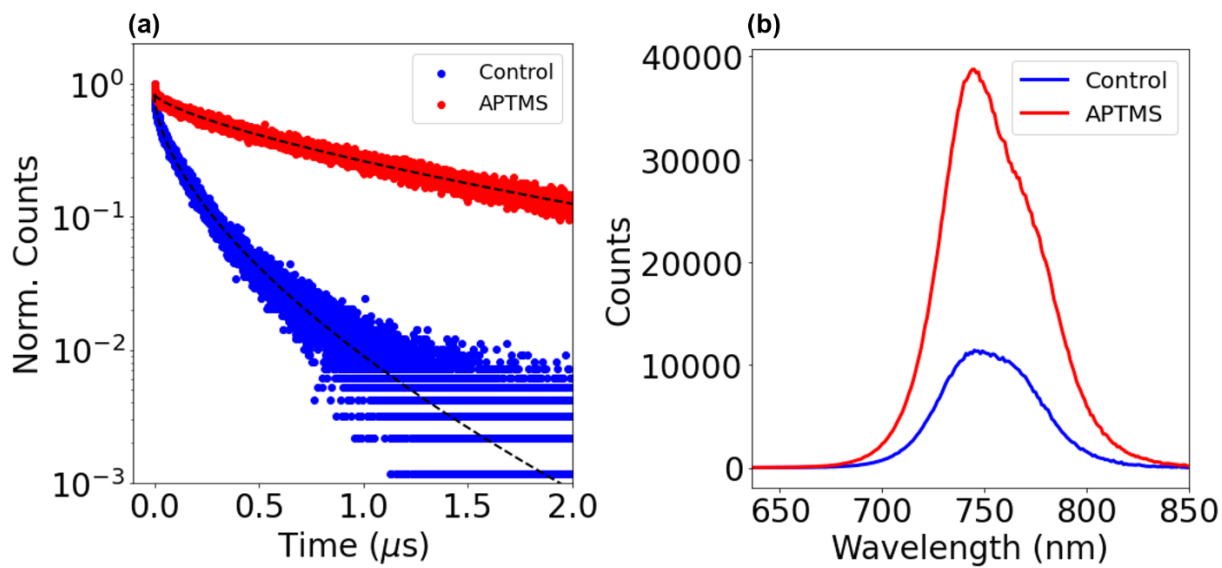

Figure 1. (a) Time-resolved PL measurements for $Cs_{0.22}FA_{0.78}Pb(I_{0.85}Br_{0.15})_3$ control and APTMS-passivated films deposited on glass substrates. Stretched exponential fits for the decay curves are shown in black. We calculated average lifetimes of 120.78 and 1015.60 ns for control and APTMS-passivated films, respectively. (b) Steady-state PL spectra for perovskite control and APTMS-passivated films deposited on glass substrates.

In order to probe ion migration in these films at the local level, we combine local electric-field poling with scanning Kelvin probe microscopy (SKPM) to measure time-dependent evolution of the surface potential following application of poling fields of both positive and negative bias. Figure 2, shows the general experimental approach, which is similar to the method Yun et al. and Richheimer et al. have used to study ion migration in unpassivated MA-based perovskites. .[33,34]

First, we perform a single pass with the cantilever to measure the topography across a single line. Next, we lift the cantilever 10 nm above the sample, apply a potential to the tip, and perform a second pass across the same line. During this pass the poling bias causes mobile charges to move towards or away from the surface, depending upon the polarity of the poling bias. Finally, we remove the poling bias, engage the Kelvin probe at the same lift height, and we measure the contact potential difference (CPD) between the tip and sample after poling. We repeat this process for every line in the image. After measuring the sample at a range of difference tip voltages, we generate a stack of CPD images in the same region. This process ensures that there is no charge injected into the film that could complicate the measurement or cause electrochemical interactions,[35] and probes the samples in the dark.

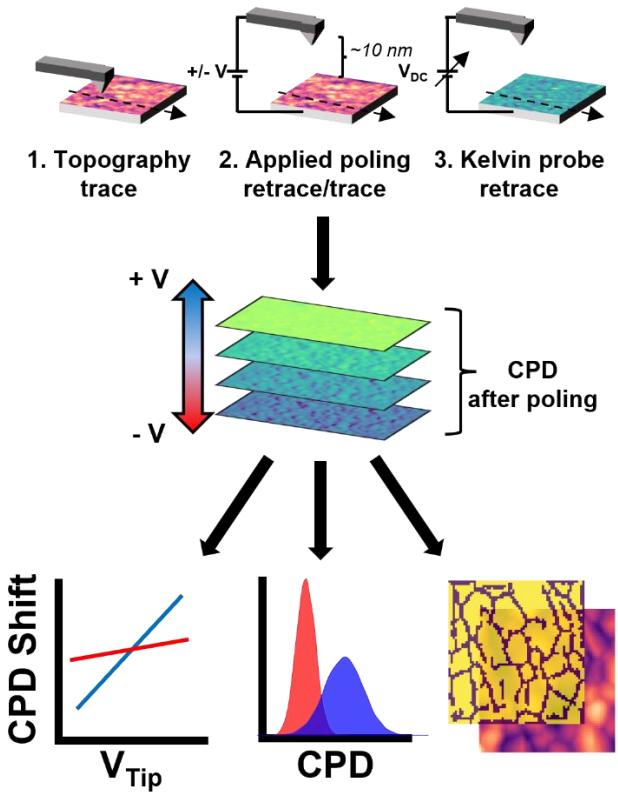

**Figure 2.** Schematic of poling-based SKPM measurement and pathways of data analysis. We measure the topography, followed by a poling step, and finally initiate a Kelvin probe loop to measure the CPD at each line in the image. The data is processed to look at the CPD shift as a function of the applied bias to the tip, the distribution in CPD to analyze heterogeneity, and differences that occur at domain interfaces relative to the domain centers.

We use this poling-based SKPM measurement to probe a $Cs_{0.22}FA_{0.78}Pb(I_{0.85}Br_{0.15})_3$ control film on ITO under biases ranging from -3 V to +3 V with steps of 1 V. Figures 3a-d shows the topography and the evolution of the CPD at applied biases of 0, +3, and -3 V. We observe a shift to more positive CPD values when applying a positive bias to the tip, which we attribute to the build up of negative charges at the surface that is consistent with accumulation of negative ionic surface charge at the film surface resulting in a vacuum level offset. We observe the opposite effect when a negative bias is applied to the tip – a large shift to negative CPD values is seen as positive charges accumulate at the surface, resulting in a vacuum-level offset of the opposite sign, which we illustrat in Figure S2. Figure 3d, measured at a tip bias of -3 V, shows a significant degree of heterogeneity in the CPD as the applied tip bias becomes more negative and net positive charge accumulates at the perovskite surface (presumably due to driving negative ions away).

Figures 3e-h shows the topography and CPD evolution of an APTMS-passivated perovskite film on ITO which reveals two key differences between the passivated and unpassivated samples. First, the surface-passivated samples show lower overall shifts in their CPD following local poling. Second, the passivated perovskite films have much more homogenous CPD distributions, both before, and after poling. Figure 3i shows the distribution of the CPD shift, which is determined by the difference between the CPD after poling and the CPD measured at 0 V. Figures j-k show the average CPD shift and full-width half max (FWHM) of the CPD shift relative to the poling bias. We see that the average CPD shift measured at the bias extremes decreases from around 100 mV to only ~20 mV. Importantly, this observation that APTMS-based passivation reduces the CPD shift induced during poling is consistent with the hypothesized suppression of ionic conductivity due to the reduction of halide vacancies. Accordingly, the FWHM of the measured CPD should reflect the extent of heterogeneity in local ion migration in a given film. Figure 3k shows that upon applying more intense negative biases, we observe a broadening of the FWHM in the unpassivated film, while the APTMS-passivated film exhibits a consistent, narrow FWHM of only ~10 mV. We propose that this difference is due to the greater heterogeneity in the control film, likely due to differences in ion mobility between the perovskite domains and the domain interfaces. In Figure S7, we show a full picture of the average CPD shift and heterogeneity as a function of poling bias for both perovskite formulations. Overall, these results show that APTMS surface passivation reduces ion migration in these wide-gap perovskite formulations.

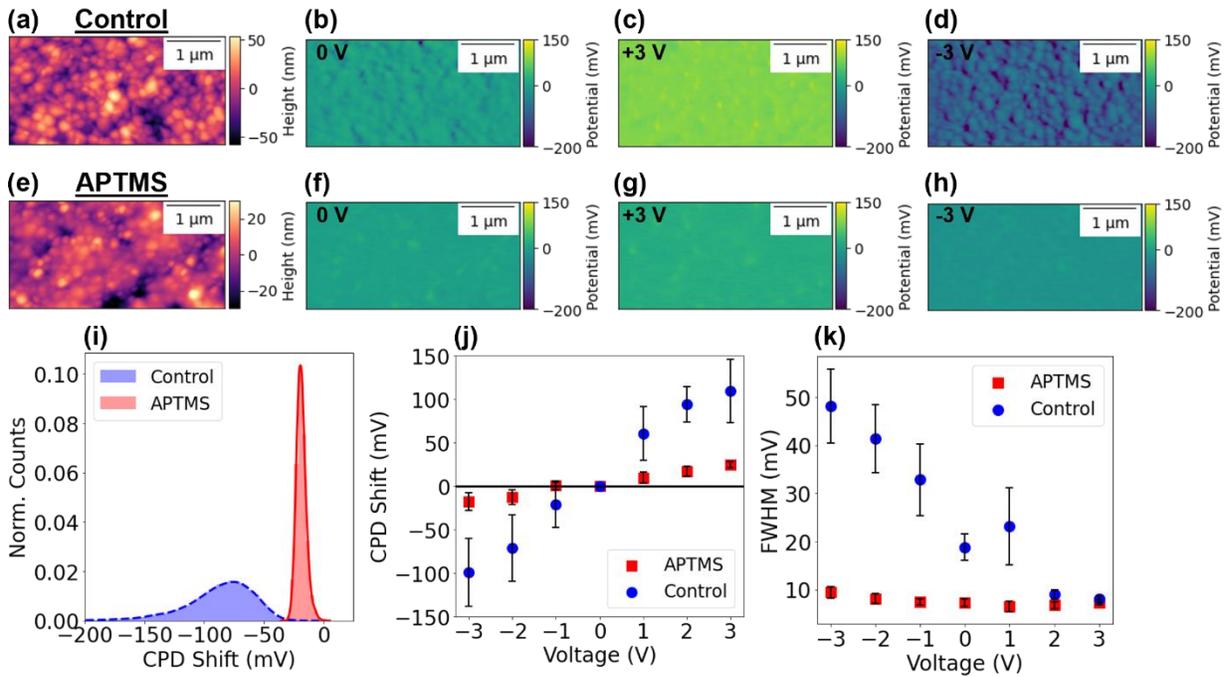

Figure 3. (a) Topography and (b-d) CPD of $Cs_{0.22}FA_{0.78}Pb(I_{0.85}Br_{0.15})_3$ control film measured with applied tip biases of 0, +3, and -3 V. (e) Topography and (f-h) CPD of APTMS passivated film measured with applied tip biases of 0, +3, and -3 V. All CPD images are shown on the same color scale to show the differences between the two samples. (i) Probability density distribution of the shift in CPD relative to the baseline CPD for control and APTMS passivated films measured at -3 V. (j) Shift in average CPD bias relative to the baseline CPD and (k) FWHM as a function of applied tip for control and APTMS passivated films. Error bars are shown as standard error of the mean for three different measurements performed on three different films each from different batches.

We recognize based on previous literature that the topographic features observed are not necessarily grain boundaries, as a single crystallite can contain multiple domains,[36,37] but for the purposes of convenience here we refer to the crystallites as "domains" and the spaces between them as "interfaces". In order to separate the differences between the perovskite domains and interfaces, we compare the mean CPDs at the domains and interfaces using several images collected at various biases of the $Cs_{0.22}FA_{0.78}Pb(I_{0.85}Br_{0.15})_3$ film. To achieve this separation, we align each CPD image based on its associated topography image and masks were manually selected to distinguish individual topographic domains. Figures 4a-b show the topography of the unpassivated perovskite film, and the mask used to separate the domains and interfaces. Using this methodology, we aggregate the CPD as a function of distance to the nearest GB pixel for control and APTMS-passivated films. Figure 4c shows the average CPD as a function of distance to the nearest domain interface for control and APTMS-passivated films measured with a -3 V bias (Figure S6 shows this calculation for all poling biases). We observe a large difference in the average CPD measured at or near domain interfaces compared to regions further away – at domain centers. In films that are surface passivated with APTMS, we see that the CPD becomes both more homogenous relative to its distance from the nearest interface, and uniform across all distances. Figure 4d shows the difference between the average CPD measured at domain centers and interfaces relative to the poling bias for unpassivated and APTMS-passivated films. We observe a negative linear trend in the CPD difference with poling biases increasing from -3 to +3 V for the unpassivated film. In contrast, we see the CPD difference remains relatively unchanged with varying biases for the passivated film. Figure S8 visualizes the difference between the domain centers and interfaces, in which we see a contrast inversion in the CPD for the unpassivated film. The larger CPD shifts at the visible interfaces are consistent with a range of literature reports suggesting increased ion motion near surface interfaces and domain interfaces.[33,35,38,39] Importantly, these new SKPM results also show that APTMS-based passivation preferentially treats domain interface-related defects, leading to significantly more homogeneous films in terms of their response to bias-induced ion motion.

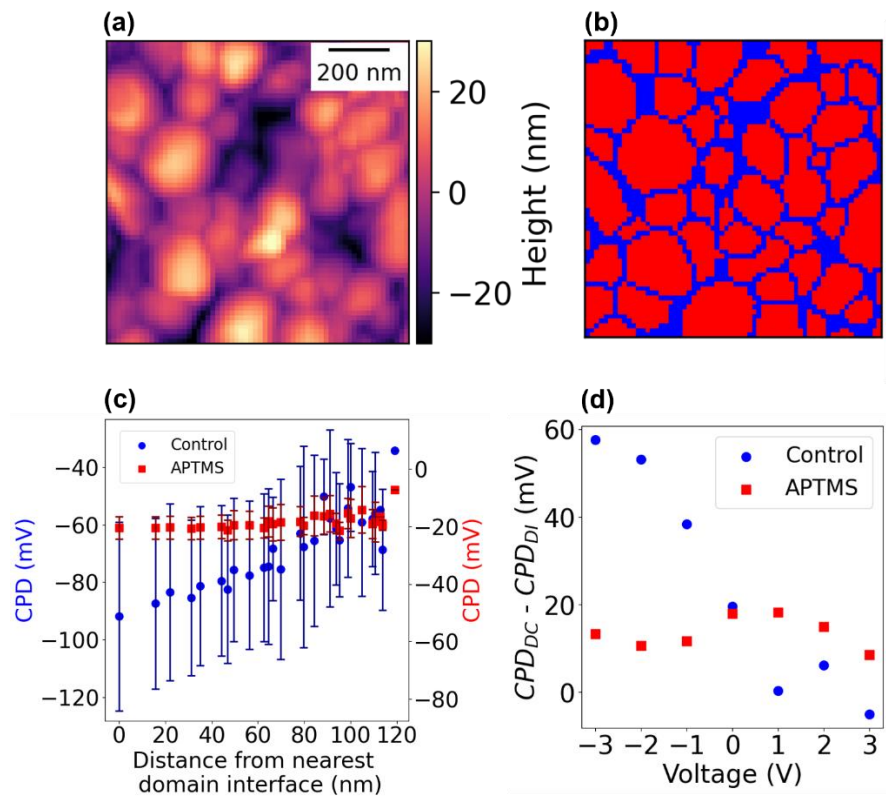

Figure 4. (a) Topography of a perovskite control film. (b) Binary mask created to separate perovskite domains (red) from domain interfaces (blue). (c) Average CPD as a function of distance from the nearest domain interface for control and APTMS-passivated films. Error bars indicate the standard deviation of the CPD probed at varying distances from nearest domain interface. (d) Difference in the CPD measured at domain centers (DC) and interfaces (DI) for control and APTMS-passivated films.

Finally, we use hyperspectral photoluminescence microscopy under laser light bias to further explore ion migration in the control and APTMS-treated films. For this measurement, we kept the perovskite samples in a dry-nitrogen environment and excited them with a 532 nm laser at 600 mW/cm$^2$. Although moderately higher than 1 Sun (100 mW/cm$^2$), we selected this illumination intensity to accelerate the effects of ion migration within a reasonable time frame. We provide further details of the measurement parameters in the Supporting Information.

Figures 5a-b show the hyperspectral photoluminescence maps of the unpassivated perovskite sample before and after light-soaking under laser light bias. Figure 5c shows the normalized photoluminescence spectra measured during the light-soaking process. Based on the time-dependent evolution of the photoluminescence, we observe a shift in the PL spectra and spatial distribution of features, which we interpret as initial phase segregation with contributions from both iodide-rich (peak emission at 780 nm) and mixed (peak emission at 740 nm) phases. This phase segregation can be visualized in the overall spectra as a shoulder peak. These regions have been attributed to an inhomogeneous elemental distribution that forms during the crystallization process.[40–42] We see that the iodide-rich regions grow significantly in size after 30 minutes of light

soaking. Figures 5d-e show the hyperspectral mapping and overall photoluminescence spectra for APTMS-passivated films undergoing light-soaking. Similar to the control film, we see initial halide segregation. In contrast to the control, we see that the growth of the iodide-rich regions is hindered by the APTMS surface passivation. In both films, we observe a slight red-shifting of the main mixed-phase emission peak, which may be attributed to demixing of the A-site cations as observed by Knight, et al.[43] Figure S9 shows the cumulative photoluminescence mapping for unpassivated and APTMS-passivated films, in which we see a consistently higher photoluminescence intensity for the surface passivated films.

We apply a wavelength threshold of 765 nm to separate the mixed and iodide-rich phases by binning the pixels based on their emission wavelength (Figure S10a), with the aim of further understanding how the mixed and iodide-rich phases evolve over the course of light-soaking. Figure S10b shows the shift in emission wavelength relative to light-soaking duration for the mixed and iodide-rich phases in control and passivation films. We can see the conversion from mixed-phase perovskite to iodide-rich perovskite during light-soaking, where regions emitting at wavelengths shorter than 765 nm, prior to prolonged light exposure, shift to longer wavelengths over time. This effect is more pronounced in the control film as compared to the APTMS-passivated film.

In Figure S10c, we determined the extent of phase segregation by calculating the fraction of iodide-rich pixels relative to mixed-phase pixels as a function of light-soaking time compared with the initial amount of phase-segregation prior to light exposure. While both the control and APTMS-passivated films show consistent red-shifting with light-soaking, the extent of phase-segregation increases to 7X its initial state after prolonged light-soaking in the control film as compared to an increase of 2X in the APTMS-passivated film. Overall, we see that APTMS passivation on these wide-gap perovskites treats defects like halide vacancies, which results increased photoluminescence quantum yields and reduce halide segregation. These photoluminescence observations are in agreement with and complement the results obtained with SKPM.

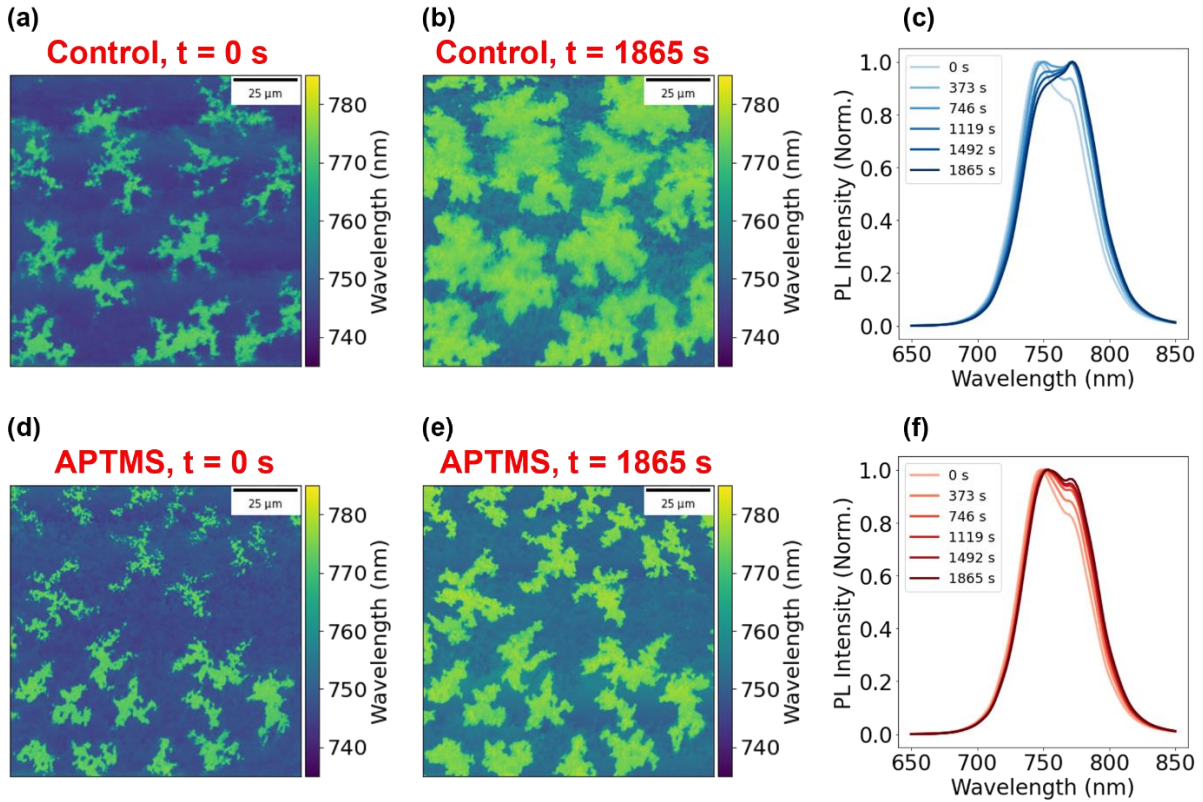

Figure 5. Hyperspectral microscopy emission images collected at light-soaking times of (a, d) 0 s and (b, e) 1865 s for (a, b) $Cs_{0.22}FA_{0.78}Pb(I_{0.85}Br_{0.15})_3$ control and (d,e) APTMS passivated films. (c,f) Histograms of the wavelength emission counts for each of the images in (a-d). Ensemble spectra for control (blue) and APTMS passivated (red) films calculated from hyperspectral maps. Samples were excited with a 532 nm laser with a fluence of ~600 mW/cm$^2$.

Studying the evolution of the surface potential with SKPM following poling reveals insight into how APTMS passivation affects ion motion in halide perovskites. Notably, we observe a significant reduction in the concentration of charges that drift from poling after surface passivation with APTMS. We examined differences in the CPD of domain centers and domain interfaces as a function of poling bias and polarity, and we observed that domain interfaces exhibit a higher amount of ion migration compared to domain centers. This difference is suppressed following passivation with APTMS. To further study the role of APTMS on ion migration, we used hyperspectral photoluminescence microscopy to explore time-dependent halide segregation during illumination of these materials. We found that halide segregation, as measured by the PL red-shift, is significantly more pronounced in unpassivated films. Taken together, these data indicate that defect passivation reduces ion migration, both in terms of domain-to-domain variations in ion migration rate as well as the overall ion migration magnitude, and that these effects are correlated with lower amounts of photoinduced halide segregation. These results highlight the importance of developing new methods to measure ion migration and provide a simple method for screening new passivating agents via AFM for beneficial ion migration properties.

Acknowledgements

This letter is based on work supported primarily by the U.S. Department of Energy (DOE-SC0013957). Part of this work was conducted with instruments supported by the University of Washington Student Technology Fee at the Molecular Analysis Facility, a National Nanotechnology Coordinated Infrastructure site at the University of Washington, which is a user facility supported in part by the National Science Foundation (Grant ECC-1542102), the Molecular Engineering & Sciences Institute, the Clean Energy Institute, and the University of Washington. We acknowledge the University of Washington Clean Energy Institute and the Washington Research Foundation. M.D.B. is supported by an NSF Graduate Student Fellowship under Grant no. DGE-2140004. D.S.G. acknowledges salary and infrastructure support from the Washington Research Foundation, the Alvin L. and Verla R. Kwiram endowment, and the B. Seymour Rabinovitch Endowment.